\documentclass{aa}  

\usepackage{graphicx}
\usepackage{txfonts}
\usepackage{lipsum}
\usepackage{subcaption}         
\usepackage{lscape}             
\usepackage{placeins}           
\usepackage{xcolor}                              
\usepackage[colorlinks=true, linkcolor=blue, citecolor=blue, urlcolor=blue]{hyperref}

\begin{document}

   \title{Gamma-ray emission from particle illumination and shock-cloud interaction in the W51 Complex
   }


\author{Alan Sunny\inst{1,2}\corrauth{alan.sunny@inaf.it}        
        \and Martina Cardillo\inst{1} \corrauth{martina.cardillo@inaf.it} 
        }
\institute{
Istituto di Astrofisica e Planetologia Spaziali -- INAF, 
Via del Fosso del Cavaliere 100, 00133 Rome, Italy
\and
Macroarea di Scienze MM.FF.NN., Università di Roma Tor Vergata,
Via della Ricerca Scientifica 1, 00133 Rome, Italy
}

   \date{Received September 30, 20XX}

 
 \abstract
  {In the current era of very-high-energy (VHE) and ultra-high-energy (UHE) $\gamma$-ray astronomy, understanding Galactic PeVatrons and their acceleration mechanisms remains a primary objective. Recent LHAASO observations of the W51 Complex make it an ideal laboratory for investigating the origin of UHE emission, particularly due to the presence of massive and dense molecular environment surrounding multiple potential particle accelerators. In this work, we study two hadronic scenarios for the W51 Complex. First, we model the direct interaction between the SNR W51C and the nearby clouds in W51B, incorporating fresh particle acceleration, shock-driven adiabatic compression, and reacceleration of permeating Galactic cosmic rays. Second, we explore an accelerator-independent illumination scenario in which the W51B cloud acts as a long-term confinement region for high-energy particles injected during an earlier epoch. We find that the direct shock-cloud interaction scenario successfully reproduces the GeV emission observed by Fermi-LAT, but fails to account for the UHE emission detected by LHAASO. In contrast, the illumination scenario naturally explains the UHE spectrum, indicating that dense molecular clouds can efficiently confine and sustain energetic hadronic populations over long timescales. Although the inferred injection history is compatible with a young SNR origin, the source-independent nature of the illumination framework also permits other accelerators within the W51 Complex. Our results therefore identify dense molecular environments as the key structures sustaining historical PeVatron activity and shaping the observed UHE $\gamma$-ray emission. 
  }


   \maketitle
   \nolinenumbers

\section{Introduction}

Supernova Remnants (SNRs) have long been considered the primary contributors to the Galactic cosmic ray (CR) population. Particle acceleration in these sources is commonly modeled within the framework of Diffusive Shock Acceleration (DSA) and the observational confirmation of CR acceleration itself comes from $\gamma$-ray detections of the characteristic “\emph{pion bump}” \citep{Blasi2013_CRorigin}. Detected by AGILE-GRID and Fermi-LAT in sources such as W44 and IC 443 \citep{giuliani2011neutral, ackermann2013detection, Giuliani2024_SNRGamma}, this spectral feature provides unambiguous proof that SNRs accelerate protons up to at least several tens of TeV.

However, despite this success at GeV energies, a major discrepancy remains: the "PeVatron problem". This refers to the historical difficulty of identifying Galactic sources capable of accelerating CRs up to the "knee" of the spectrum ($\sim 10^{15}$ eV) \citep{cristofari2021hunt, vink2022nonthermal}. While young SNRs such as Tycho exhibit $\gamma$-ray spectra extending up to low TeV, they often display cutoffs below the PeV scale. Conversely, middle-aged remnants like W44 and IC 443 show spectral breaks at much lower energies ($10$ - $100$ GeV), indicating a lack of CRs at TeV and higher energies \citep{vink2022sources}. More recently, LHAASO has reported TeV emission from the W44 region, although its spectral and morphological properties are still under investigation \citep{LHAASO2025_W44}. This energy gap is critical because the detection of emission exceeding 100 TeV acts as a 'smoking gun' for the presence of accelerated particles at or near the PeV scale.  While the \emph{pion bump} confirms the existence of hadronic CRs, only the observation of these highest-energy photons can determine if a source is currently or was historically a PeVatron, thus bridging the gap between GeV-scale observations and the known Galactic CR spectrum \citep{cristofari2021hunt, cao2024does}. Furthermore, theoretical expectations suggest that PeV hadronic acceleration in SNRs should occur predominantly within the first $\sim 100$ years after the explosion \citep{cardillo2015_CRTypeII}, yet observations of young remnants like Cassiopeia A place stringent limits on their UHE proton energy budget, challenging their efficiency as long-lived PeVatrons \citep{cao2024does}.

The standard view considered $\gamma$-ray emission above 100 TeV with a hard spectrum as a ‘smoking gun’ for hadronic acceleration. This was because leptonic scenarios are generally expected to exhibit much steeper spectrum at those energies due to Klein-Nishina decay. However, the landscape of this problem changed significantly with the recent detection of ultra-high-energy (UHE; $\gtrsim$ 100 TeV) $\gamma$-ray emission by LHAASO. These observations revealed several sources including young massive star clusters (YMSCs) \citep{lhaaso2021ultrahigh}, microquasars (MQs) \cite{Cao2025_lhaasoMQ} and pulsar wind nebulae (PWNe) \cite{lhaaso2021_crabpeta}, majorly leptonic accelerators, redefining the PeVatron problem.  But the most surprising results came from SNRs: several UHE emitting sources detected by LHAASO are associated with old or middle aged ($\sim15$-$30$ kyr) SNRs that cannot reach PeV energies alone, requiring a more sophisticated look.

Historically, the presence of dense ambient material has been a leading explanation for enhanced SNR emission. Specifically, when an SNR shock directly interacts with a nearby molecular cloud (MC), it involves rich physics including fresh particle acceleration at transmitted shocks, compression of molecular material, and reacceleration of pre-existing Galactic CRs. A detailed treatment of this was presented in \citet{cardillo2016supernova}, who developed a self-consistent framework to explain the GeV emission in W44 \citep{uchiyama2012_fermiW44, cardillo2014_AGILE_W44, abe2025_W44MagicFermi}. The most recent shift in this narrative is the illumination scenario, which provides a natural framework for explaining UHE emission even from older remnants. In this picture, escpaed particles subsequently interact with nearby dense MCs, producing delayed hadronic $\gamma$-ray emission \citep{Gabici_MC_Illumination2009, gabici2019origin}.

In recent years, the illumination scenario has been extensively explored and has emerged as a leading explanation for several LHAASO detected sources. \citep{mitchell2024lhaaso, mitchell2024exploring, kar22, sarkar24}. In this picture, the UHE $\gamma$-ray emission observed today does not originate from the presently evolved and weakened SNR shock itself, but rather from high-energy particles that escaped the accelerator at earlier stages and later interacted with a nearby dense MC. At the same time, the accelerator responsible for injecting these particles does not necessarily need to be a SNR. Other energetic environments, such as star-forming regions or stellar winds, may also inject particles capable of illuminating dense MC structures. The viability of this scenario fundamentally depends on the presence of a sufficiently dense cloud capable of efficiently confining the injected particle population and enhancing particle interactions. This motivates exploring a source-independent illumination model that focuses on the MC as the primary site of particle confinement and delayed $\gamma$-ray production, without assuming a specific accelerator. Our aim is to determine whether the observed properties of the cloud alone are sufficient to sustain the injected particle population and reproduce the measured emission. MCs associated with powerful particle accelerators therefore provide ideal laboratories for testing this scenario.

Within this context, the W51 Complex stands out as an ideal target. Among the LHAASO sources, W51 is particularly compelling due to strong evidence for efficient CR acceleration by the SNR W51C \citep{jogler2016revealing, aleksic2012morphological, 2009HESS}, along with clear signatures of direct interaction between the SNR shock and the adjacent MCs in W51B \citep{Brogan_2013OHMaser, tu2025molecular}. In addition, the complex hosts stellar clusters and a candidate PWNe \citep{Ginsburg_2015W51Clouds, XMM_2014CXO}, providing multiple potential acceleration sites. W51B itself is a massive star-forming region located in close proximity to the SNR and contains roughly 10-30\% of the total mass of the whole W51 Complex \citep{Ginsburg_2015W51Clouds}, making it dense enough to effectively confine CRs. These properties make W51 an excellent laboratory to investigate both shock-cloud interactions and the indirect illumination of confined CRs within MCs as drivers of $\gamma$-ray emission, as discussed in the following sections in detail.

The paper is arranged as follows: in the Sec \ref{Sec:W51} we give a brief review on the W51 Complex; in Sec. \ref{Sec:Theory} we introduce our theoretical setup and methodology on both direct interaction and source-independent illumination; in Sec. \ref{Sec:Results} we discuss our results on how the UHE emission could be explained with illumination and discuss the contribution from direct acceleration of CRs, adiabatic compression of clouds and reacceleration of permeating Galactic CRs; and our conclusions are drawn in Sec. \ref{Sec:Conclusion}.

\section{The W51 Complex}\label{Sec:W51}
The recent detection by LHAASO of $\gamma$-ray emission extending up to $\sim 200$ TeV from the W51 Complex \citep{cao2024evidence} represents a significant milestone for Galactic PeVatron studies. W51 is a massive MC complex with a diameter of $\sim 100$ pc and a total mass of $\sim 10^6\,M_{\odot}$, making it one of the most active star-forming regions in the Galaxy. The system hosts multiple potential particle accelerators and is commonly divided into three main components: the star-forming regions W51A and W51B, and the middle-aged SNR W51C (G49.2-0.7). W51C has a shell-type morphology in radio wavelengths \citep{koo1995rosat, THOR_2016} and has an estimated age of $\sim$18-30 kyr. Located at a distance of $\sim 5.5$ kpc, it has a radius of $\sim 24$ pc and an inferred kinetic energy of up to $3.6 \times 10^{51}$ erg, highlighting its potential role as an efficient accelerator of high-energy particles.

Prior to the LHAASO observations, the W51 Complex, particularly W51C, had been extensively studied in $\gamma$-rays by Fermi-LAT \citep{abdo2009_fermi, jogler2016revealing} and  MAGIC \citep{aleksic2012morphological} with additional reports from H.E.S.S. \citep{2009HESS}, Milagro \citep{2009milagro}, and HAWC \citep{20203HAWC}, covering an energy range from $\sim 50$ MeV to $\sim 30$ TeV. One of the most compelling results came from Fermi-LAT, which reported a characteristic \emph{pion-bump} like feature in the $\gamma$-ray spectrum of W51C, providing strong evidence for a hadronic origin of the emission \citep{jogler2016revealing}. Despite this, the older age of W51C makes it difficult for the remnant alone to account for the UHE $\gamma$-ray emission observed at energies approaching 200 TeV, as standard DSA is expected to reach PeV energies only during the early evolutionary stages of SNRs. 

The recent LHAASO observations reveal an extended TeV source associated with the W51C-B region. The emission is well described by a power law with exponential cutoff (PLExpCut) with a spectral index of $2.48 \pm 0.08$ and an angular extension of $0.17^\circ \pm 0.02^\circ$ \cite{cao2024evidence}. These values are broadly consistent with the spectral indices and source extensions previously reported by MAGIC \citep{aleksic2012morphological} and Fermi-LAT \citep{abdo2009_fermi, jogler2016revealing}: $2.58 \pm 0.07$ and $2.50 \pm 0.18$ and the extensions of $0.12^\circ \pm 0.02^\circ$ and $\sim 0.22^\circ$, respectively.

W51B is not considered to be an active star-forming region but it contains, as discussed above,  10-30\% mass of the complex and hosts three stellar clusters, observed both in radio and far-infrared and spatially coincident with the $\gamma$-ray extended emission: G48.9-0.3, G49.2-0.3. and G49.0-0.3. Two of these, G48.9-0.3 and G49.2-0.3 are potential PeV accelerators \citep{Pirola2026_w51,cao2024evidence}. On the other hand W51A is characterized by ongoing and vigorous high-mass star formation \citep{Fujita2021_SFRW51A}. It contains two massive protocluster candidates, W51 Main and W51 IRS 2. The W51 Main region is estimated to have formed only a small fraction of its potential and is expected to efficiently produce another $10^{4} \sim 10^{6} \text{M}_{\odot}$ of stars \citep{Ginsburg_2015W51Clouds}. But it should be noted that there is no strong spatial coincidence with the UHE emission reported by LHAASO here. 

Finally the complex hosts the PWNe candidate CXO J192318.5+140305 \citep{Chandra_2005CXO, XMM_2014CXO}. We know relatively little about this source especially in $\gamma$-rays, although MAGIC has suggested a possible association, indicating that it could also act as a potential accelerator \citep{aleksic2012morphological}. More precisely, morphological analysis from MAGIC indicates that the VHE emission's centroid and extension are largely energy-independent between 300 GeV and 5 TeV, with the primary emission maximum firmly coinciding with the shocked gas at the interface of W51C-B. Although a tail-like feature extending toward the south-eastern PWNe candidate becomes more prominent above 1 TeV, its relative contribution is estimated at only $\sim20$ \% (ref Fig.4 in \citet{aleksic2012morphological}). It is also important to note that \citet{cao2024evidence} reports an enhanced flux below 10 TeV, which is approximately 1.8 times higher than that measured by MAGIC \citep{aleksic2012morphological}. While this discrepancy may be partially explained by differences in detector sensitivity and angular resolution between extensive air shower arrays and imaging atmospheric Cherenkov telescopes (IACTs) \citep{cao2024evidence}, its origin is still not fully understood.

A leading interpretation of the observed $\gamma$-ray emission invokes interactions between CRs accelerated at the W51C shock and dense molecular material in the nearby W51B region. This picture is well supported by multi-wavelength observations: LHAASO $\gamma$-ray maps show enhanced emission concentrated around the W51C-W51B region, while independent studies at other wavelengths report clear signatures of SNR-MC interaction, including chemical evidence for shock-processed molecular clumps \citep{tu2025molecular} and OH (1720 MHz) maser emission in the overlapping C-B region \citep{zhang2017disentangling}. In particular, observations of 1720 MHz OH masers and CO(3-2) emission reveal a ring of shocked gas partially encircling the non-thermal radio source W51B\_NT \citep{Brogan_2013OHMaser}. More recently, \citet{tu2025molecular} investigated the shock-processed molecular clumps in “Clump 2” \citep{koo1997binteraction}, mapping an  $82.5'' \times82.5'' $  region (corresponding to $\sim 2.2\, \text{pc} \times 2.2\, \text{pc}$ at a distance of 5.5 kpc), revealing re-formed molecular gas behind a dissociative J-shock.

Motivated by this, a direct shock-cloud interaction naturally emerges as a promising mechanism for the high-energy $\gamma$-ray emission. However, within standard DSA expectations, even detailed modeling of the interaction zone suggests that this scenario alone is insufficient to fully account for the observed TeV flux, pointing to the need for an additional component. In our previous work \citep{sunny2025studying}, the UHE emission could be reproduced under extreme assumptions, requiring unrealistically strong magnetic fields and idealized conditions. Here, we adopt more realistic parameters consistent with recent studies.

We therefore adopt W51 as a representative system and investigate both direct SNR-MC interaction and CR illumination scenarios. It is also important to note that alternative models have explored the retention of particles near SNRs due to strongly suppressed diffusion in the circumstellar medium, which can result in a significant population of non-confined particles residing in the shock precursor region \citep[e.g.][]{Celli2019_exploringEsc,Schure2014_CRtogalPrecursor}. While this mechanism creates a 'halo' of particles that can produce a broken-power-law $\gamma$-ray spectrum \citet{Celli2019_exploringEsc}, for an SNR at the age of W51C (18-30 kyr), this approach generally predicts spectral breaks at much lower energies than the UHE emission detected by LHAASO, suggesting that the UHE component is unlikely to originate from the current particle population at the shock. 

At the same time, recent studies have proposed that the UHE emission in some LHAASO sources, including W51, can be explained within an SNR-MC interaction framework \citep{xian2025investigating}. So we explore this interaction framework, but incorporating detailed shock-cloud modeling, including shock-driven adiabatic compression and the reacceleration of pre-existing Galactic CRs. In addition, motivated by the illumination framework discussed above, we explore for the first time a source-independent CR illumination scenario for the W51 Complex, where particles injected during earlier epochs diffuse into and remain confined within the dense molecular environment of W51B.


\begin{figure*}
    \centering
    \includegraphics[width=0.8\textwidth]{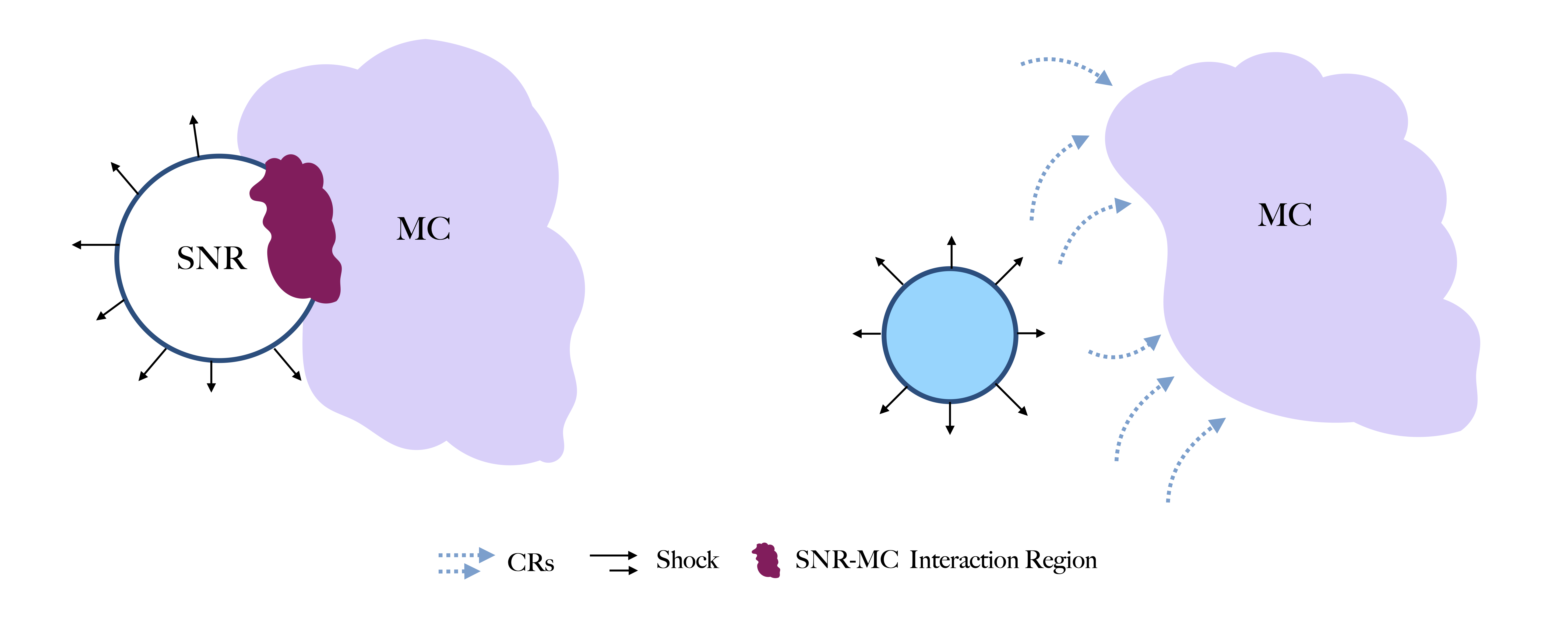}
    \caption{Schematic of the two particle interaction scenarios considered for the W51 complex: direct interaction between the W51C SNR shock and the adjacent MC in W51B, resulting in shock compression within the interaction region (left); and past illumination, where CRs escaping from a nearby accelerator, for example a younger phase of W51C (illustrated by the blue circle), diffuse into the surrounding MC (right).}
    \label{fig:theory_graphic}
\end{figure*}

\section{Methodology} \label{Sec:Theory}
The complete theoretical framework considered in this work is illustrated in Fig.~\ref{fig:theory_graphic} and can be broadly divided into two scenarios: (a) current-time interaction dominated by in-situ processes, including acceleration of fresh particles at the shock, compression-induced emission from shock-crushed clouds, and reacceleration of pre-existing Galactic CRs, and (b) past illumination of the W51B MC by high-energy particles, which may originate from the early evolutionary stages of the SNR or from other nearby accelerators,

\subsection{Shock-cloud direct interaction}

As discussed above, shock-cloud interaction is expected to be the dominant contributor to the observed $\gamma$-ray emission from this region, particularly in light of the morphological and spectral results reported by MAGIC \citep{aleksic2012morphological} and Fermi-LAT \citep{jogler2016revealing}. Our primary analysis therefore focuses on the SNR–MC interaction region to investigate the extent to which this mechanism can account for the observed VHE to UHE $\gamma$-ray emission. Earlier \citet{blandford1982radio} introduced the “crushed cloud” scenario, where a supernova-driven shock compresses a dense cloud, increasing the emission and explaining the radio observations. If the shock velocity and ambient density are sufficiently high, the post-shock gas becomes radiative, emitting ionizing radiation throughout the shock’s path. Initially, the gas is compressed, and as recombination sets in, further radiative cooling enhances compression and increases density. This mechanism was later explored in the context of $\gamma$-ray emission from W44 by \citet{cardillo2016supernova}.

For a shock velocity ($v_{\text{sh}}$) of $\sim 400$ km/s in W51C, the formation of a radiative zone requires a minimum column density $N_{\text{cool}} \approx 3 \times 10^{17} v_{\text{sh}}^4\, \text{cm}^{-2}$, leading to a minimum cloud density $n_{cloud} \gtrsim n_{cool}$. 
Also, the compression factor between the downstream of the shock (density $n_d$) and the crushed cloud (density $n_{cc}$), $s \equiv \frac{n_{cc}}{n_d}$ (or compression due to radiative cooling), can be limited by magnetic or thermal pressure. The compression due to radiative cooling enhances the particle energy spectrum normalization by a factor of $s^{2/3}$, and the momentum of each particle as $p \rightarrow p\,s^{1/3}$. Detailed formalism is given in \citet{cardillo2016supernova}

\subsubsection{Acceleration and shock induced cloud compression} \label{sec:acc+comp}
The primary mechanism for $\gamma$-ray emission in SNRs is the radiation of particles accelerated at the shock front. Both hadronic and leptonic components are generally modeled with a power-law momentum distribution:
\begin{equation}\label{eq:Particle_PL}
    f_{pr,el}(p) = k_{pr,el} \left( \frac{p}{p_{\text{inj}}} \right)^{-\alpha},
\end{equation}
for protons and electrons, given by the indices $pr$ and $el$, respectively, with a spectral index $\alpha$ and injection momentum $p_{\text{inj}}$. The normalization $k_i$ is set by the CR acceleration efficiency $\xi_{\text{CR}}$. See Equations 9-11 in \citet{cardillo2016supernova} for further details.

In this framework, particles already accelerated at the shock front undergo additional energization through compression within the interaction region. This process may contribute to the enhanced flux measured by LHAASO around $\sim 10$ TeV compared to MAGIC \citep{aleksic2012morphological}. We note, however, that part of the discrepancy between LHAASO and MAGIC has been attributed to differences in instrumental response, angular resolution and sensitivity to diffuse extended emission \citep{cao2024evidence, lhaaso2021_crabpeta}. Nevertheless, we investigate these processes to provide a more detailed physical description of the region. We define the compression factor as:
\begin{align}
s \equiv \frac{n_{\text{cc}}}{n_d} = \frac{n_{\text{cc}}}{r_{\text{sh}} n_0},
\end{align}
where $n_0$ the ambient density and $r_{sh} = \frac{n_d}{n_0}$ is the compression ratio at the shock. Following the framework of \citet{blandford1982radio}, we assume that in SNR W51C the gas compression induced by cooling is constrained by magnetic pressure. Given $B_0 = b \sqrt{n_0/\text{cm}^{-3}}\, \mu\text{G}$, the increase in the compressed magnetic field and from it the density of the crushed cloud are estimated as:
\begin{equation}
\frac{B_{cc}^2}{8 \pi} = n_0\mu_H v_{\text{sh}}^2 \, \rightarrow \, n_{cc} \simeq 94\, \left(\frac{n_0}{\text{cm}^{-3}}\right)^{3/2}\, \left(\frac{B_0}{\mu\,\text{G}}\right)^{-1}\, \left(\frac{v_{\text{sh}}}{10^7\,\text{cm/s}}\right),
\end{equation}
Here, the parameter $b$ depends on the Alfvén velocity, defined as $b = v_{A}/1.84, \text{km s}^{-1}$. Its typical value is of order unity in the ISM, while in MCs it generally ranges between $0.3$ and $3$. Now due to compression, the resulting particle spectrum then shifts: 

\begin{equation}
  f^{\prime}(p) = f(s^{-1/3} p),
\end{equation} 

As highlighted by Sushch and Borse \citep{sushch2023limits}, previous models may overestimate the amount of material contained within the compressed cloud. They impose a strict upper limit on the total number of particles, constrained by the volume swept by the shock:
\begin{equation}\label{eq:SushchBorseUL}
N_{\text{max}} = V_{\text{SNR}}\, n_0,
\end{equation}
Introducing a volume filling factor $\chi$ (defined as $V_{\text{cc}}/V_{\text{SNR}}$), the volume of the crushed cloud can be written as $V_{\text{cc}} = \chi V_{\text{SNR}}$, where $V_{\text{SNR}}$ denotes the remnant volume. The total number of particles within the crushed cloud is then $N_{\text{cc}} = V_{\text{cc}}\, n_{\text{cc}} = \chi\, V_{\text{SNR}}\, n_{\text{cc}}$. By expressing the cloud density in terms of a compression factor $s$, this becomes $N_{\text{cc}} = \chi\, V_{\text{SNR}}\, s\, n_0$. Consistency with Eq. \ref{eq:SushchBorseUL} therefore requires $\chi s \leq 1$.

In practice, the limited interaction area and finite duration of the shock-cloud interaction imply $\chi s \ll 1$, with a more realistic upper bound of $\chi s \leq 0.1$ (see \citet{sushch2023limits} for details). Although this constraint excludes some parameter regimes explored in earlier works \citep{tutone2021multiple}, where such limits were not enforced, we adopt it here to ensure a more conservative and physically consistent treatment. Our general approach follows the methodology outlined in \citet{cardillo2019orion, cardillo2016supernova}, and we will fix parameters known from observational and detailed magneto-hydro-dynamic simulations and vary the other unknown parameters considering physical reasoning and the fit. 

\subsubsection{Reacceleration of Galactic CRs}

As discussed earlier, molecular line studies indicate the presence of fast, radiative J-type shocks propagating through dense molecular material, characterized by strong gas compression and rapid molecular re-formation \citep{tu2025molecular}. Such environments typically favor freshly accelerated particles at the SNR blast wave. At the same time we have a second particle population, namely pre-existing Galactic CRs, in addition to shock-accelerated particles. These Galactic CRs may undergo compression and possible reacceleration within the same shock-cloud interaction region.

However, \citet{sushch2023limits} argue that compressed Galactic CRs alone are unlikely to produce sufficient $\gamma$-ray emission, and that reacceleration primarily shifts the spectrum in energy rather than significantly altering its normalization. Despite this, even a subdominant contribution is expected to yield some radiative output, particularly due to enhanced confinement of Galactic CRs within the cloud.

Previous work by \citet{cardillo2016supernova} has demonstrated that reacceleration can contribute to the emission observed by AGILE-GRID. Motivated by this, we adopt a similar framework to investigate its role in the W51C-B region. In this context, we explore whether the combined effects of reacceleration and adiabatic compression can enhance the Galactic CR contribution, especially in the GeV regime. Although this involves a distinct particle population, these processes occur within the same physical environment as the freshly accelerated particles discussed earlier. We therefore follow the formalism and Galactic CR spectrum presented in \citet{cardillo2016supernova}, adopting the same simulation setup for the interaction region as described in Section~\ref{sec:acc+comp}.

\subsection{Particle illumination}
\label{subsec:Part_Illum}

If the MCs are sufficiently dense and the diffusion timescale satisfies $t_{D} \geq age$, where $age$ represents the particle evolution timescale, then particles from a past epoch can remain effectively confined within the MCs without escaping. In such a scenario, they continue to evolve inside the cloud and produce delayed $\gamma$-ray emission that we can see now. This is what we refer to as the "illumination" scenario.

A key requirement of the illumination scenario is the presence of a sufficiently powerful particle accelerator in the vicinity of the MC. The W51 complex is a particularly rich environment, hosting several sources capable of accelerating particles to very high energies. The SNR W51C is a plausible contributor considering the proximity to the MC, it may have injected UHE particles into the MC with minimal propagation losses in a relatively early evolutionary stage. The region also contains stellar clusters in W51B (G48.9-0.3, G49.2-0.3. and G49.0-0.3), and the PWNe candidate CXO J192318.5+140305, all located within the extended LHAASO emission region along with massive star-forming region and associated stellar winds in W51A.

Rather than focusing on a specific accelerator, our objective is to investigate the role of the MC itself. Motivated by the spatial association between the extended $\gamma$-ray emission and the W51B giant MC (GMC) region, we examine whether the cloud can efficiently confine an injected CR population and produce delayed hadronic $\gamma$-ray emission. This source-independent approach allows us to isolate the effects of particle confinement within the cloud without introducing the large and highly degenerate parameter space associated with modeling individual accelerator classes and their escape histories. We therefore begin directly with particle injection into the MC.

Using the observed properties of the W51B GMC, we constrain the particle energetics required to reproduce the measured $\gamma$-ray emission. Since the candidate sources are all expected to inject predominantly hadronic CRs, we adopt an electron-to-proton luminosity ratio of 0.01. The resulting emission is then calculated to assess whether the cloud alone can account for the observed UHE $\gamma$, independent of the nature of the accelerator.


To model the W51B GMC, we approximate it as a uniform cloud with averaged physical parameters. From \citet{Ginsburg_2015W51Clouds}, W51B spans Galactic longitudes $48.8^\circ \leq \ell \leq 49.4^\circ$ and latitudes $-0.5^\circ \leq b \leq -0.1^\circ$, corresponding to an angular extension of approximately $0.3^\circ$. Assuming a distance of 5.5 kpc, this translates to a physical size of $\sim 30$ pc. With an estimated mass of $\sim 10^5 \, \mathrm{M}_\odot$, we derive an average density of $\sim 120 \, \mathrm{cm}^{-3}$, consistent with \citet{Pirola2026_w51}. 

The simulation in this work is performed using \textsc{GAMERA} \citep{GAMERA_2022Hahn}, an open-source library designed to simulate the time-dependent evolution of non-thermal leptonic and hadronic populations. It enables the construction of sophisticated radiation models by solving the transport equations within a customizable astrophysical framework, allowing for a significantly more realistic treatment of particle spectra than steady-state approximations. In this framework, we numerically solve the time-dependent transport equation given by:
\begin{equation}\label{Eq:Transport}
\frac{\partial N}{\partial t} = Q(E, t) - \frac{\partial}{\partial E} \left[ b(E,t), N(E,t) \right] - \frac{N(E,t)}{t_{D}},
\end{equation}
where $N(E,t)$ represents the differential particle distribution as a function of energy and time. The term $Q(E,t)$ denotes the injection spectrum of particles, while $b(E,t)$ describes the total energy loss rate, incorporating all relevant radiative and non-radiative processes. The final term accounts for particle escape from the system, parameterized through the characteristic diffusion timescale $t_D$.

The diffusion timescale encapsulates the energy-dependent escape of particles from the emission region and is defined following the formalism of \citet{kar22} as:
\begin{equation}\label{eq:tdiff_main}
t_D = \frac{R^2}{D(E)},
\end{equation}
where R is the characteristic spatial scale of the system, in this case the height of the cloud. The diffusion is modeled as,
\begin{equation}
D(E) = D_0 \left(\frac{E}{E_0}\right)^{\delta},
\end{equation}
where we take $D_0 = 10^{26} cm^2/s$ considering a better particle confinement, $E_0 = 4\,\text{GeV}$ is the reference energy, and $\delta = 0.33$, corresponding to a Kolmogorov-type turbulence spectrum \citep{kar22,Strong04_Diffuse}. The adopted coefficient is consistent with the suppressed diffusion expected in dense molecular environments and near CR accelerators \citep{cui2016_youngillum,Gabici2010_constraintsslowdiff}. This prescription naturally accounts for the energy dependence of particle diffusion within the MC.

Finally, as discussed before, the expected extension of W51B is $\sim 30$ pc. So we define $t_D$ from Eq \ref{eq:tdiff_main} as:
\begin{equation}
t_{D(pr,el)} = \frac{2.46 \times 10^6}{\left(E_{pr,el}/E_0\right)^\delta} \,\,\mathrm{yrs},
\label{eq:tdiff_W51}
\end{equation}
where $pr$, $el$ represent protons and electrons, respectively. 

As mentioned above, we have constraints on MC parameters, on the contrary we try to constrain the time of injection, the age of particle evolution within the MCs and the injection spectrum of the particles, which we will discuss in Sec. \ref{Subsec:Results_illum}.

\section{Results}\label{Sec:Results}
\begin{figure*}[h!]
\centering
\includegraphics[width=0.48\textwidth]{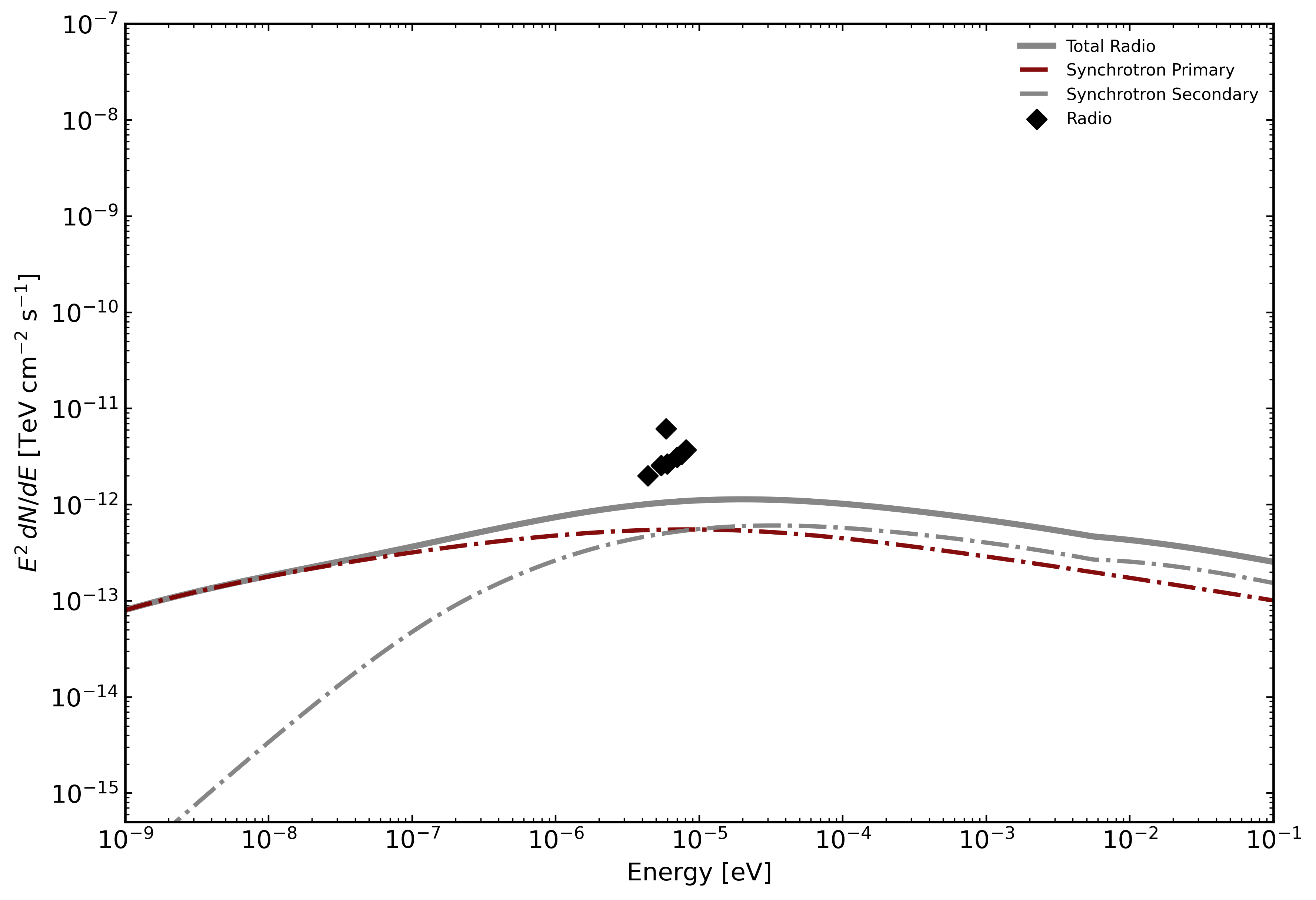}\hfill
\includegraphics[width=0.48\textwidth]{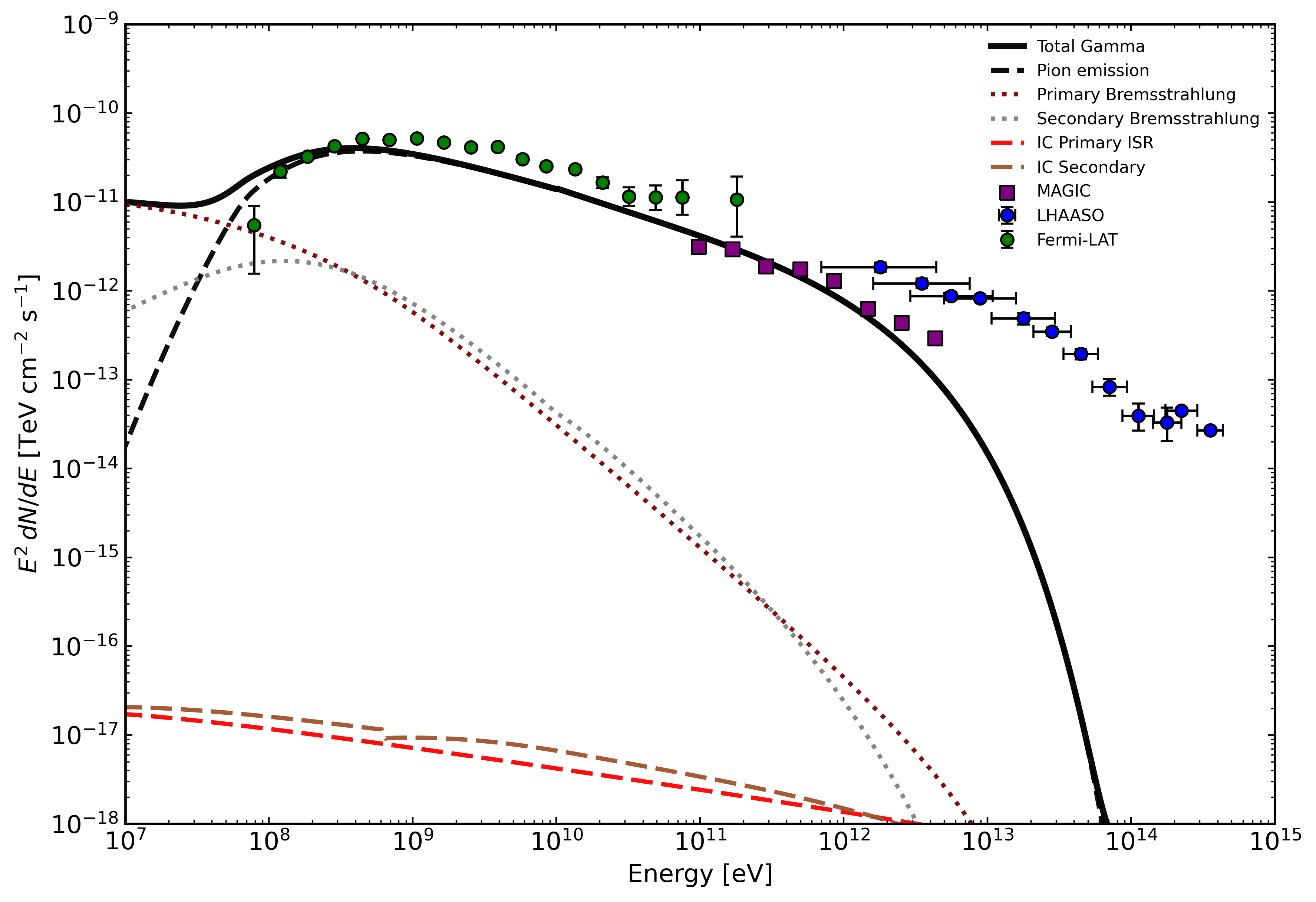}
\caption{Radio spectral energy distribution of the W51C-B region (left) and $\gamma$-ray spectral energy distribution of the interaction region (right).
The $\gamma$-ray data are from \emph{Fermi-LAT},
\emph{MAGIC}, and
\emph{LHAASO}. Radio flux measurements are extracted from the VLA THOR Galactic survey for the region encompassing the W51C-B interaction. This fit is attained by considering an SNR of age 18 kyrs interacting with the medium for $\sim 11$ kyrs. We consider a shock velocity of 360 km/s propagating into a medium of an initial ambient density $n_{0} = 20\, \text{cm}^{-3}$. A filling factor of 0.01 (1\%) and a Kolmogorov perturbation spectrum with $ k_T = 2/3$ is considered. The emission components include pion decay (thick black dashed line), as well as primary and secondary contributions from bremsstrahlung (dotted lines), inverse Compton scattering (dashed lines), and synchrotron radiation (dot–dashed lines).}
\label{fig:Acc-CC-W51}
\end{figure*}

\begin{figure*}[h!]
\centering
\includegraphics[width=0.48\textwidth]{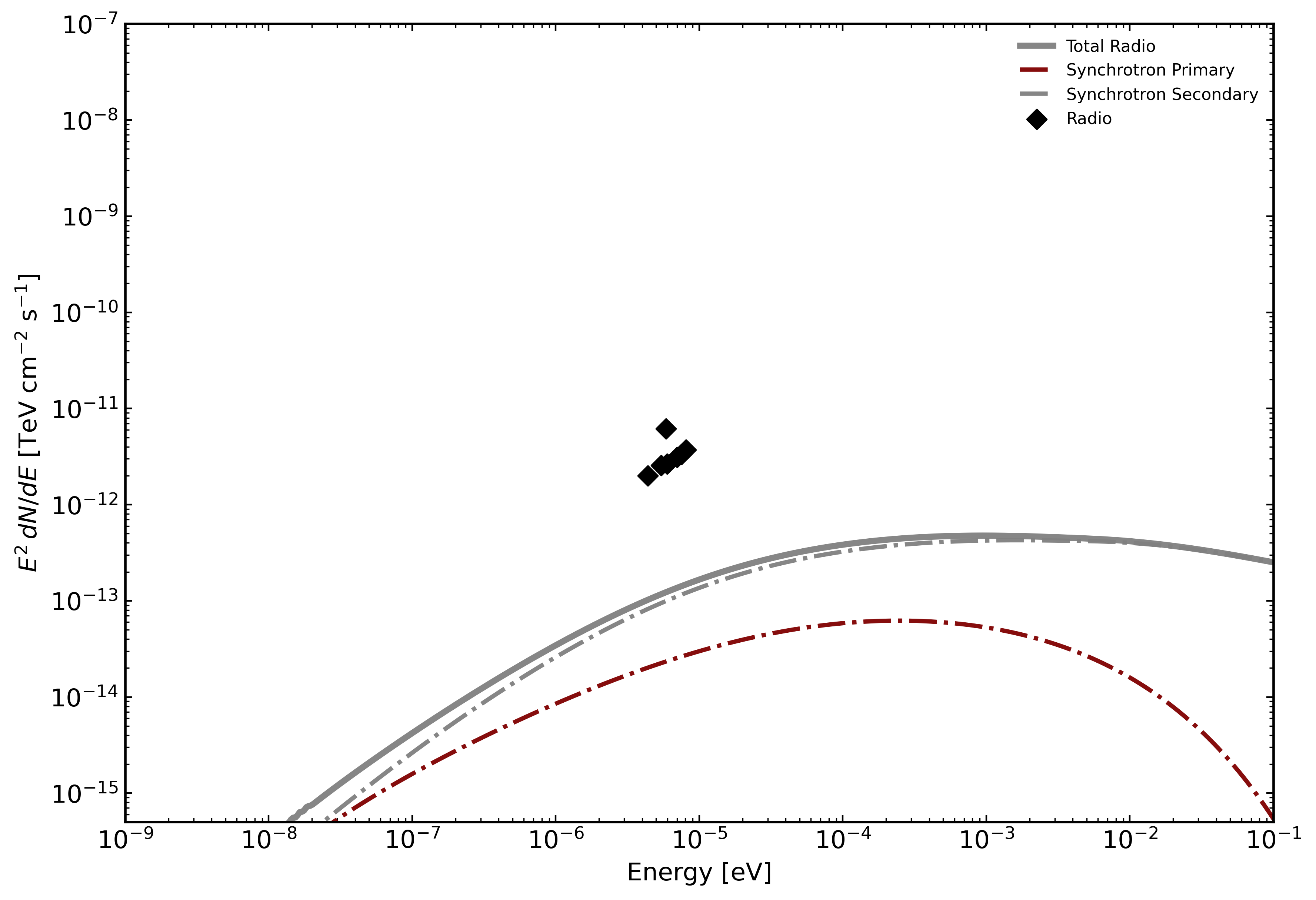}\hfill
\includegraphics[width=0.48\textwidth]{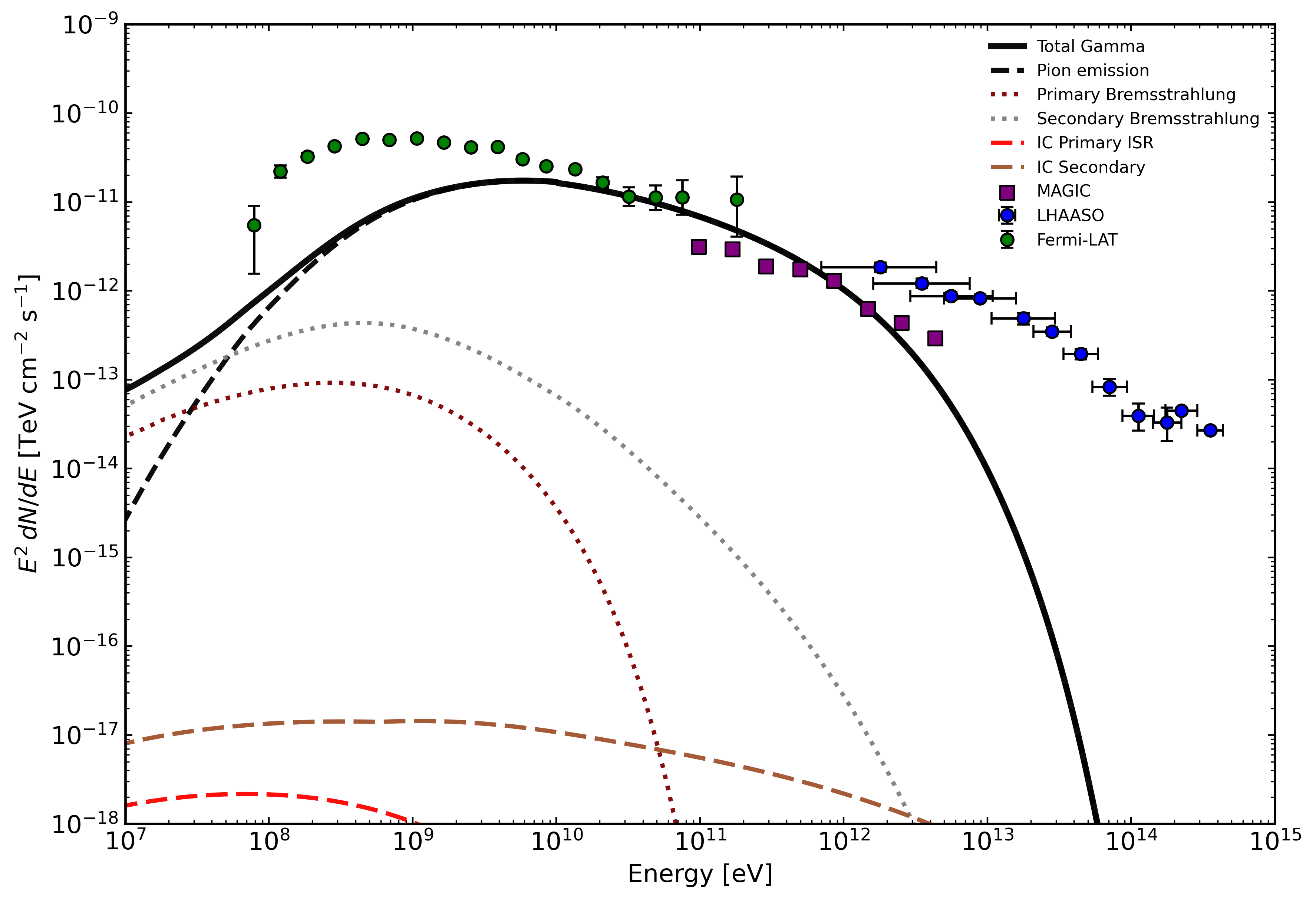}
\caption{
Multiwavelength spectral energy distribution of the W51 region considering initial contributions from Galactic CR reacceleration and compression of MCs: radio emission (left) and $\gamma$-ray emission (right). The model adopts the same physical parameters as in the acceleration + compression case shown in the Fig \ref{fig:Acc-CC-W51}, as both scenarios describe the same interaction region. The particle population is based on the Voyager 1 local interstellar spectra, following \citet{cardillo2016supernova} and as characterized by \citet{Potgieter2014_VoyagerLISM}.
}
\label{fig:Reacc-CC-W51}
\end{figure*}
In this section, we present our results by considering the two origins of emission: freshly accelerated particles that undergo compression at the shock-cloud interaction region, with additional contributions from reacceleration and compression of pre-existing Galactic CRs and particles illuminating the GMC (with averaged conditions) of W51B (Fig. \ref{fig:theory_graphic}). These results are compared with $\gamma$-ray observations from LHAASO \citep{cao2024evidence}, Fermi-LAT \citep{jogler2016revealing}, and MAGIC \citep{aleksic2012morphological}, along with radio measurements from the VLA THOR survey \citep{THOR_2016, THOR_2020} for the interested region. 

For the illumination scenario, we include contributions from pp interactions, Bremsstrahlung, inverse Compton (IC), and synchrotron emission. As discussed earlier, this scenario is modeled within a proton-dominated framework. We neglect the contribution from secondary particles (e.g. electrons produced in hadronic interactions) because their radiative output is expected to be subdominant compared to the dominant pion-decay $\gamma$-ray emission from diffusive protons illuminating the cloud \citep{IllumDiffProtons_2012}.

In contrast, for the direct interaction scenario, we include emission from pion decay as well as both primary and secondary contributions to Bremsstrahlung, IC, and synchrotron processes. This is motivated by the effects of shock compression, which enhance the gas density and magnetic field, thereby increasing both the production rate and radiative efficiency of secondary particles \citep{cardillo2016supernova}.

We first examine the direct interaction scenario, assessing its ability to reproduce the data and its limitations, before introducing the illumination scenario as a complementary mechanism to account for the UHE component.

\subsection{Direct interaction: acceleration, compression and Galactic CR reacceleration}

\begin{figure*}[h!] 
  \centering
  \includegraphics[width=\textwidth]{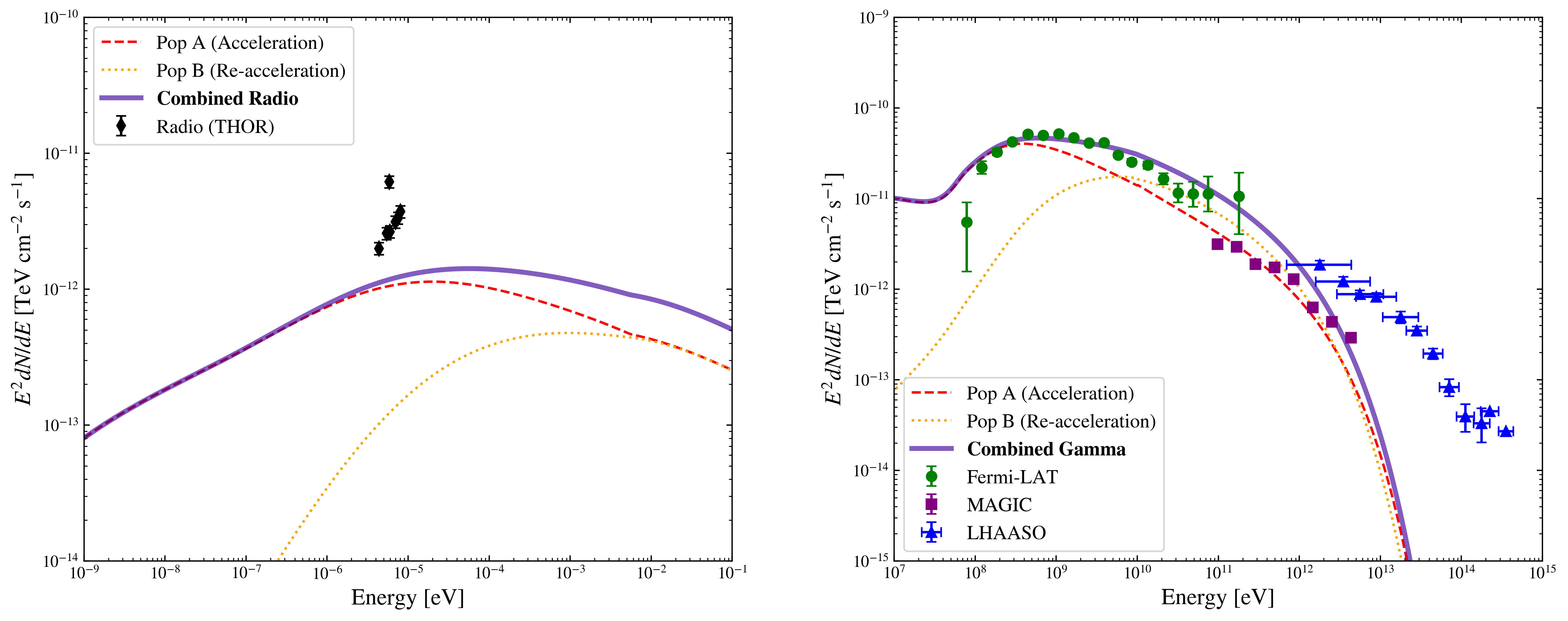}
  \caption{Combined spectral energy distribution of the W51 SNR-MC interaction region. The total emission is modeled as the sum of two particle populations: (A) freshly accelerated particles at the interaction site (red dashed line; for details refer Fig. \ref{fig:Acc-CC-W51}), and (B) the per-existing Galactic CR population (yellow dotted lines; for details refer Fig. \ref{fig:Reacc-CC-W51}). Left: Radio emission. Right: $\gamma$-ray emission.}
  \label{fig:Joint_SED}
\end{figure*}

Guided by observational and theoretical constraints, we start by fixing key parameters in our model: radius of the SNR $R_{\mathrm{SN}} = 24\,\text{pc}$ \citep{cao2024evidence}, and local ambient density of $n_0 = 20$-$25\,\text{cm}^{-3}$ \citep{fang2010non}. Note that the ambient density here is the local density at the MC interaction site and is different from what we consider in the illumination scenario which is the averaged density of the whole MC. For W51C, theoretical estimates suggest a shock velocity of $v_{\mathrm{sh}} \sim 400\,\mathrm{km/s}$, consistent with the Sedov evolutionary stage and broadly compatible with previously reported values of $480\text{-}490\,\mathrm{km/s}$ \citep{koo1995rosat}. In our modeling, we adopt a shock velocity of $360\,\mathrm{km/s}$, representative of the large-scale SNR blast wave expanding into a comparatively low-density medium. When such a fast shock encounters nearby dense gas, it is expected to drive a slower transmitted shock within the cloud, leading to strong radiative compression. Observational studies indeed indicate transmitted J-type shocks with velocities of $\sim 70 $ - $ 100 \,\,\text{km/s}$ propagating into molecular material, as inferred from H I kinematics \citep{koo1997interaction} and molecular chemistry analyses \citep{tu2025molecular}. While our code does not explicitly follow the time-evolution of the transmitted shock velocity inside the cloud, the resulting post-shock gas density obtained in our simulations ($\sim 2.2 \times 10^{3}$ cm$^{-3}$ for an initial density of $\sim 20$ cm$^{-3}$) is consistent with high compressed densities inferred from these independent studies. 

As for the magnetic field, we adopt a value of $b=3$, as constrained  by the data. this value lies within the range expected for MCs \citep{cardillo2016supernova}, implying an ambient magnetic field in the interaction zone, of $B_0 = 14 \mu \mathrm{G}$. Although there is some debate in the literature regarding the age and distance of W51, we adopt the values from \citet{lhaaso2021ultrahigh, tian2013high, zhang2017disentangling}, taking the distance $d = 5.5\, \text{kpc}$ and the age as $t_{age} = 18,000$ years. Following \citet{sushch2023limits}, we use a filling factor of 0.01 (1\%). We safely assume the acceleration efficiency of the SNR to be $1.5 \times10^{-3}$, indicating that only a very small fraction of the shock energy is converted into non-thermal particles, consistent with a localized and inefficient acceleration scenario such as middle aged SNR shock-cloud interaction. A Kolmogorov perturbation spectrum with $ k_T = 2/3$ is assumed. 

The remaining free parameters, such as the correlation length $L_c$ and the interaction timescale $t_{\mathrm{int}}$, are varied and constrained based on the best-fit results. We assume that the interaction with the MC has been ongoing for $\sim 11{,}000$ years, shorter than the SNR age $(t_{\mathrm{age}} = 18{,}000\,\mathrm{yr})$.

The correlation length in the interstellar medium is typically of the order $L_c \sim 10$ -$100$ pc, corresponding to the scale at which turbulence is injected, likely by supernova activity. For MCs this can be much smaller, $L_c \sim 0.01$ - $0.1$ pc. In our model, we adopt a value of $L_c = 3 \times 10^{17}\,\mathrm{cm}$, which lies well within plausible limits. Finally, we define the particle distribution by assuming a simple power-law form (Eq.~\ref{eq:Particle_PL}), with a spectral index of 4.4 and an electron-to-proton ratio of $\sim 10^{-2}$. The resulting spectrum, accounting for the joint contributions of particle acceleration and adiabatic compression, is illustrated in Fig. \ref{fig:Acc-CC-W51}.

While the combined contributions from particle acceleration and adiabatic compression align well with the MAGIC data, the resulting normalization is insufficient to reproduce the observed Fermi-LAT and LHAASO flux. Furthermore, it fails to provide a consistent fit across the radio emission levels. Consequently, we extend the simulation to isolate the individual contributions of reacceleration and compression of Galactic CRs. This secondary component operates under the same physical conditions as the accelerated particles, differing only in the initial particle parameters. For the seed Galactic CR population, we adopt the spectrum derived from Voyager data \citep{Potgieter2014_VoyagerLISM} as described in \citet{cardillo2016supernova}, assuming an injection index of 4.5, a value consistent with theoretical expectations. Although more recent observations from both Voyager 1 and 2 have refined the interstellar spectrum at the lowest energies \citep{Cummings2016_Voyager, Stone2019_Voyager2}, the 2014 parametrization remains a robust and widely accepted baseline. It accurately captures the unmodulated spectral slope required to quantify the 'seed' population available for reacceleration and compression within the SNR-MC system. The final resulting emission spectrum from this reacceleration phase is shown in Fig. \ref{fig:Reacc-CC-W51}.

As expected the model is not able to reach the UHE regime and remains significantly sub-dominant in the radio band, but it is noteworthy that the combined contribution from reacceleration and compression reaches the required normalization to accurately reproduce several Fermi-LAT and MAGIC data points. This suggests that the presence of strong shocks in an SNR of this class can indeed facilitate a substantial contribution from Galactic CR reacceleration and compression.

Although the $\gamma$-ray emission originates from two distinct particle populations, freshly accelerated particles and the ambient Galactic CR pool, the resulting $\gamma$-rays are expected to emerge from the same spatial region. Consequently, we perform a joint fit incorporating both components to provide a comprehensive model of the source. The resulting multi-wavelength spectral energy distribution is presented in Fig. \ref{fig:Joint_SED}.

This joint model incorporating both direct particle acceleration and the reacceleration of ambient Galactic CR populations, along with adiabatic compression, effectively reproduces the Fermi-LAT flux and some of the MAGIC data points. However, this integrated framework remains insufficient to explain the UHE emission detected by LHAASO. Based on these results, we conclude that while the SNR-MC direct interaction regime provides a robust explanation for the GeV to low-TeV emission, consistent with previous studies, it fails to account for the UHE regime. This result is not surprising taking into account the theoretical framework for which SNR can reach those energies only in the very early phase of their life.

We also want to note that earlier studies, such as \citet{jogler2016revealing}, successfully reproduce the Fermi-LAT and MAGIC data using a broken power-law proton spectrum. Their approach assumes constant injection, simplified cooling, and a single-zone framework, testing hadronic (pp) as well as leptonic (Bremsstrahlung and IC) processes. While such phenomenological models are effective in describing the observed spectral shape, they do not explicitly incorporate the underlying physical processes governing particle acceleration and interaction. In contrast, our model is physically motivated, taking into account the shock physics and the shock-cloud interaction region in much detail.

We can also see that SNR-MC interaction is not sufficient to fit the observed radio emission. The inability of our direct interaction model to reproduce the total observed radio flux (see Fig.~\ref{fig:Joint_SED}) arises from its focus on the W51C-B interaction region. Previous studies by \citet{fang2010non} have shown that, in middle-aged SNRs such as W51C, the bulk of the radio synchrotron emission originates from the part of the shock expanding into the ISM with high Mach numbers and amplified magnetic fields. In contrast, the hadronic $\gamma$-ray emission is primarily associated with the shock-cloud interaction region, the focus of this work. Moreover, \citet{sushch2023limits} showed that the compression of ambient Galactic electrons contributes only a small fraction ($\sim$ 10\%) of the observed radio luminosity in middle-aged SNRs. Therefore, the underprediction of the global radio emission by our SNR-MC interaction model is consistent with these theoretical expectations, suggesting that the radio component is unlikely to originate predominantly from the direct interaction region. Given that the emission is measured from the broader W51C-B complex, alternative contributions become plausible, including stellar clusters in W51B, or the PWNe candidate.

\subsection{Indirect interaction: Illumination by escaped CRs}\label{Subsec:Results_illum}

As discussed previously, the illumination scenario is modeled using the \textsc{GAMERA} framework \citep{GAMERA_2022Hahn}. We begin by fixing the well-constrained physical parameters of the system. As discussed in Sec. \ref{subsec:Part_Illum}, the averaged ambient density is limited to $\sim 120\,\mathrm{cm}^{-3}$; in our simulations, we adopt a safe value of $n_{MC} = 105\,\mathrm{cm}^{-3}$. In this density regime $(\lesssim 300\,\mathrm{cm}^{-3})$, the magnetic field strength exhibits significant dispersion and is typically of the order of $5\sim10\,\mu\mathrm{G}$ \citep{Crutcher_2012MagField}. We therefore assume an average MC magnetic field $B_{MC} = 7\,\mu\mathrm{G}$. This value is lower than the one assumed by LHAASO, $B_{MC}=50 \mu\mathrm{G}$, based on the non-thermal radio emission of the entire SNR shell  \citep{cao2024evidence}. This is a global simplification because for the specific UHE emission region associated with the W51B GMC, this value is likely unrepresentative. Their interpretation relies on a steady-state approach, whereas \textsc{GAMERA} allows for a time-dependent solution to the transport equation. 
We also wish to stress that the magnetic field here ($B_{MC}$) is for the GMC and is different from what we consider in the direct interaction scenario, which is the ambient magnetic field ($B_0$) at the interaction zone. The remaining parameters are constrained through spectral fitting, primarily guided by the UHE data.

\begin{figure}[h!] 
  \centering
  \includegraphics[width=0.5\textwidth]{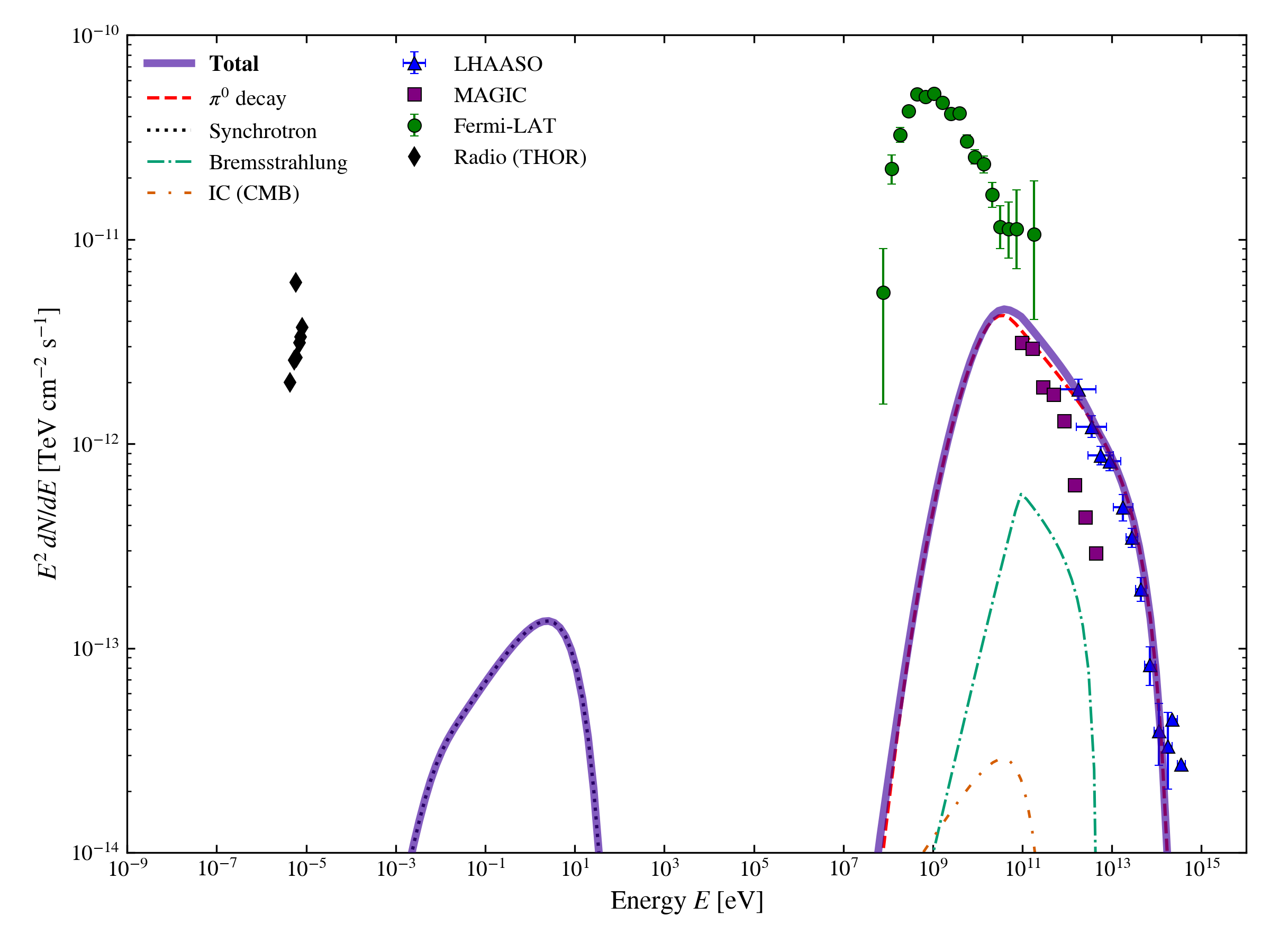}
  \caption{Total emission from the W51B GMC resulting from the illumination scenario. The cloud is modeled with an average density of $n_{MC} = 105\, \mathrm{cm}^{-3}$ and a magnetic field strength of $B_{MC} = 7\, \mu\mathrm{G}$. The parameters of the injected particle spectrum are listed in Table~\ref{tab:Inj_Particle}, with particles injected into the cloud over a duration of 420 years. The resulting emission spectrum is computed after evolving the particle population within the MC for 18 kyr.}
  \label{fig:Illum_SED}
\end{figure}

The emission spectrum shown in Fig.~\ref{fig:Illum_SED} is obtained using the particle injection parameters, at the GMC, listed in Table \ref{tab:Inj_Particle}, assuming a simple power-law injection. This choice is motivated by the source-independent nature of our framework, which seeks to evaluate the confinement potential of the GMC without being tied to the specific escape history of any single source. The resulting spectrum exhibits a high-energy cutoff around $\sim 20$ TeV consistent with the underlying parent proton population and its propagation-dependent energy losses.

\begin{table}[h!]
\caption{Parameters of injected particle spectrum into the GMC that achieved the best fit given in Fig. \ref{fig:Illum_SED}}
\label{tab:Inj_Particle}
\centering
\begin{tabular}{ccc}
\hline\hline
Parameters & Protons & Electrons \\
\hline
Maximum injected energy (TeV) & 400 & 10 \\
Minimum injected energy (TeV) & $10^{-1}$ & $10^{-1}$ \\
Luminosity (erg/s) & $4.5 \times 10^{38}$ & $4.5 \times 10^{36}$  \\
Spectral index & 2.3 & 2.3 \\
\hline
\end{tabular}
\end{table}

A key set of parameters to constrain in the illumination scenario are the relevant timescales. The energy-dependent diffusion timescale within the MC is defined in Eq.~\ref{eq:tdiff_W51}. In addition, we introduce the particle injection time into the cloud, $t_{\mathrm{inj}}$, and the system age, $age$, defined in Sec. \ref{subsec:Part_Illum}, both constrained through spectral fitting.

An important result is that reproducing the characteristic LHAASO spectrum, using the injected particle population listed in Table~\ref{tab:Inj_Particle}, requires a relatively large system age. However, this age cannot be arbitrarily large, as excessive evolution would reduce the $\gamma$-ray normalization due to pp interactions. At the same time, the condition $t_D > age$ must be satisfied to ensure particle confinement within the cloud. We find that $age \approx 18{,}000$ years provides an optimal fit to the data. Comparison of the diffusion time scales for both the particles with respect to $age$ is briefly discussed in Appendix \ref{App:A}. This is also interesting because the particle evolution timescale of 18 kyr coincides with the current estimated age of the SNR.

For this choice of $age$ and the particle parameters listed above, an injection timescale of $t_{\mathrm{inj}} \sim 420$ years is required to reproduce the observed spectrum. Given the adopted luminosities, this corresponds to a total injected energy of $\sim 10^{49}$ erg, which is well within the expected energetics of a SNR. 


The inferred injection timescale ($t_{\mathrm{inj}} \sim 420$ yr) can be associated with the early evolutionary phase of a SNR, when particle acceleration is expected to be most efficient and capable of reaching PeV energies. Within the SNR interpretation, however, the energetics of the injected particle population demands a careful look. As discussed above, our model requires the injection of protons with energies up to $\sim 400$ TeV at the location of the MC. If the progenitor explosion associated with W51C occurred at an offset from the dense cloud environment, particles accelerated during the early PeVatron phase would need to propagate through the complex ambient medium before reaching the cloud, inevitably undergoing energy losses during transport. A self-consistent treatment of these propagation and loss processes would require a much better understanding of the W51B environment and the evolutionary state of W51C, likely involving detailed shock-evolution and magneto-hydro-dynamic simulations, which is not the aim of this work. Similar treatment would also apply to any accelerator candidate located within the region, further motivating our accelerator-independent approach.  Nevertheless, from our source-independent model, we can conclude that a hadronic accelerator injecting particles into a dense environment where they are efficiently confined can successfully explain the UHE emission of W51 Complex observed by LHAASO mainly through pion-decay resulting from pp interactions.

A preliminary Markov Chain Monte Carlo (MCMC) consistency check of the illumination model parameter space (Appendix \ref{App:B}) shows broad agreement with our physically motivated solution. However, the computational cost of the \textsc{GAMERA} particle evolution calculations prevents a fully converged analysis at present. We therefore defer a complete MCMC study to future work.

\section{Discussions and Conclusions} \label{Sec:Conclusion}

We are currently in the era of VHE-UHE $\gamma$-ray astronomy, where understanding particle acceleration up to PeV energies remains a central challenge. While the exact acceleration mechanisms responsible for these $\gamma$-ray emitters are still under active debate, investigating their surrounding environments provides a promising avenue for progress. In this context, recent observations of the W51 Complex by LHAASO \citep{cao2024evidence} offer a valuable opportunity to probe the origin of UHE emission from candidate PeVatron sources. Earlier detections of the characteristic \emph{pion bump} like feature by Fermi-LAT \citep{jogler2016revealing} have established the SNR W51C as a hadronic accelerator at GeV energies. However, the detection of $\gamma$-rays extending beyond 100 TeV necessitates a re-examination of standard accelerators and acceleration scenarios.  

Previous studies of the $\gamma$-ray emission from the W51 complex \citep{jogler2016revealing,xian2025investigating,Pirola2026_w51,cao2024evidence} have largely focused on direct acceleration scenarios and broadband phenomenological interpretations. In this work, we extend the direct SNR-MC interaction framework by incorporating shock-driven adiabatic compression and reacceleration of pre-existing Galactic CRs. In addition, we investigate a source-independent illumination scenario, which has not previously been explored specifically for the W51 Complex. In this scenario, particles injected during earlier epochs propagate into and remain confined within the dense molecular environment of W51B, producing the observed $\gamma$-ray emission.

We model the direct interaction scenario following the formalism of \citet{blandford1982radio, cardillo2016supernova}, considering two particle populations: freshly accelerated particles at the shock and the pre-existing Galactic CR pool. As the shock propagates into nearby MCs, particle acceleration is accompanied by shock-induced adiabatic compression of the cloud material, which plays a crucial role in shaping the non-thermal emission. In the “crushed cloud” picture introduced by \citet{blandford1982radio}, the interaction of the SNR shock with dense gas leads to strong compression, further enhanced by radiative cooling in the post-shock region. This results in an increase in density and magnetic field strength, boosting the overall emission. This framework was extended to $\gamma$-ray production in systems such as W44 by \citet{cardillo2016supernova}.

In this work, both accelerated and reaccelerated particle populations are subject to compression. Using the Voyager 1 CR spectrum from \citet{Potgieter2014_VoyagerLISM}, we quantify the contribution of re-accelerated Galactic CRs alongside the freshly accelerated component. We find that, although acceleration and reacceleration along with compression considered individually cannot reproduce the observed flux levels, their combined contribution can account for the GeV emission reported by Fermi-LAT as shown in Fig. \ref{fig:Joint_SED}, consistent with the observed morphology overlapping the W51C-B region \citep{jogler2016revealing}. However, even within this framework, the interaction scenario governed by the SNR fails to explain the UHE emission detected by LHAASO. Given the extended nature of the emission, spanning both the W51C and W51B regions beyond the immediate interaction zone, we therefore explore an “illumination” scenario in which particles injected years before propagate into the surrounding GMC and produce delayed $\gamma$-ray emission.

In this paper we consider a parameter space which represents the averaged properties of the W51B GMC derived from already known observations and simulations. The best agreement with the observational data was obtained considering particles injected into the cloud for 420 years with the spectrum parameters given in Table \ref{tab:Inj_Particle}. The subsequent evolution of particles is modeled using energy-dependent diffusion (Eq. \ref{eq:tdiff_W51}) over a timescale of $\sim$ 18 kyr, yielding the final spectral energy distribution shown in Fig. \ref{fig:Illum_SED}. The inferred short injection duration could point towards the early stages of SNR evolution, when acceleration is expected to be most efficient \citep{Bell2004_turbulent,Ptuskin2005_Spectrum}. 
More generally this injection model starts from the particle injection at the MCs and the particle spectrum can be considered source independent. While a young SNR origin remains a viable interpretation, alternative sources within the complex, such as particles from star-forming region, or stellar winds, cannot be excluded. Nevertheless, the key result is robust: \emph{a hadronic accelerator injecting particles into a dense environment, combined with efficient confinement within the cloud, can naturally reproduce the observed UHE emission from the W51 Complex as detected by LHAASO through pion-decay resulting from pp interactions.} It is also important to note that neither of the models considered in this study can reproduce the observed radio emission. This mismatch strongly suggests that the radio-emitting population is physically distinct from the population responsible for the GeV-UHE $\gamma$-ray emission.

A key next step is to more tightly constrain the parameter space using dedicated MCMC approaches, as outlined above, which we defer to future work. At present, the interpretation remains inherently complex, largely due to the extended nature of the emission and the limited angular resolution and point spread function of LHAASO. A more definitive picture will require better resolved observational data, particularly from next-generation $\gamma$-ray facilities such as the Cherenkov Telescope Array Observatory (CTAO) \citep{Hofmann2024_CTAO}, and ASTRI Mini-Array \citep{Scuderi2022_ASTRI, Vercellone2024_ASTRI}. In particular, the improved angular resolution, sensitivity, and point spread function of CTAO and ASTRI Mini-Array will enable a much more detailed spatial characterization of the emission across the W51 Complex. Given that our model predicts a relatively compact GeV component alongside a more extended UHE emission tracing the W51B MC environment, future observations will be crucial in disentangling these components and providing a decisive test of the proposed scenario.

\begin{acknowledgements}
      We thank Dr. Jagdish Joshi and Dr. Abhijit Roy for valuable discussions on the illumination models, and Dr. Antonio Tutone for his support  during the initial stages of this work on the direct interaction model. We also acknowledge Prof. Marco Tavani for his support. AS is financially supported by PNRR - CTA+ PROGRAM (Proposal: IR0000012) PhD fellowship, funded by the European Union - NextGenerationEU and approved by the MUR.
\end{acknowledgements}

\bibliographystyle{bibtex/aa}
\bibliography{references}

\begin{appendix}

\onecolumn

\section{Diffusion time vs the evolution age of particles in the MC}
\label{App:A}
Here we present the particle diffusion time scale, from Eq. \ref{eq:tdiff_W51} vs the energy of the injected particles along with the total age of evolution of the particles inside the MC. It is given for both proton (bold line) and electron (grey dashed-lines) populations, post injection into the MC, considering the injection spectrum given in Table \ref{tab:Inj_Particle}.  We can see that for the full particle energy range for both protons and electrons, the $t_D > age$, that is 18000 years (red dotted line). This tells us that the injected particles are well confined within the MC without escaping considering 18 kyrs. The high-energy particles will only start to escape from the MCs when the $age > 50$ kyrs. 
\begin{figure}[h!]
    \centering
    \includegraphics[width=0.6\linewidth]{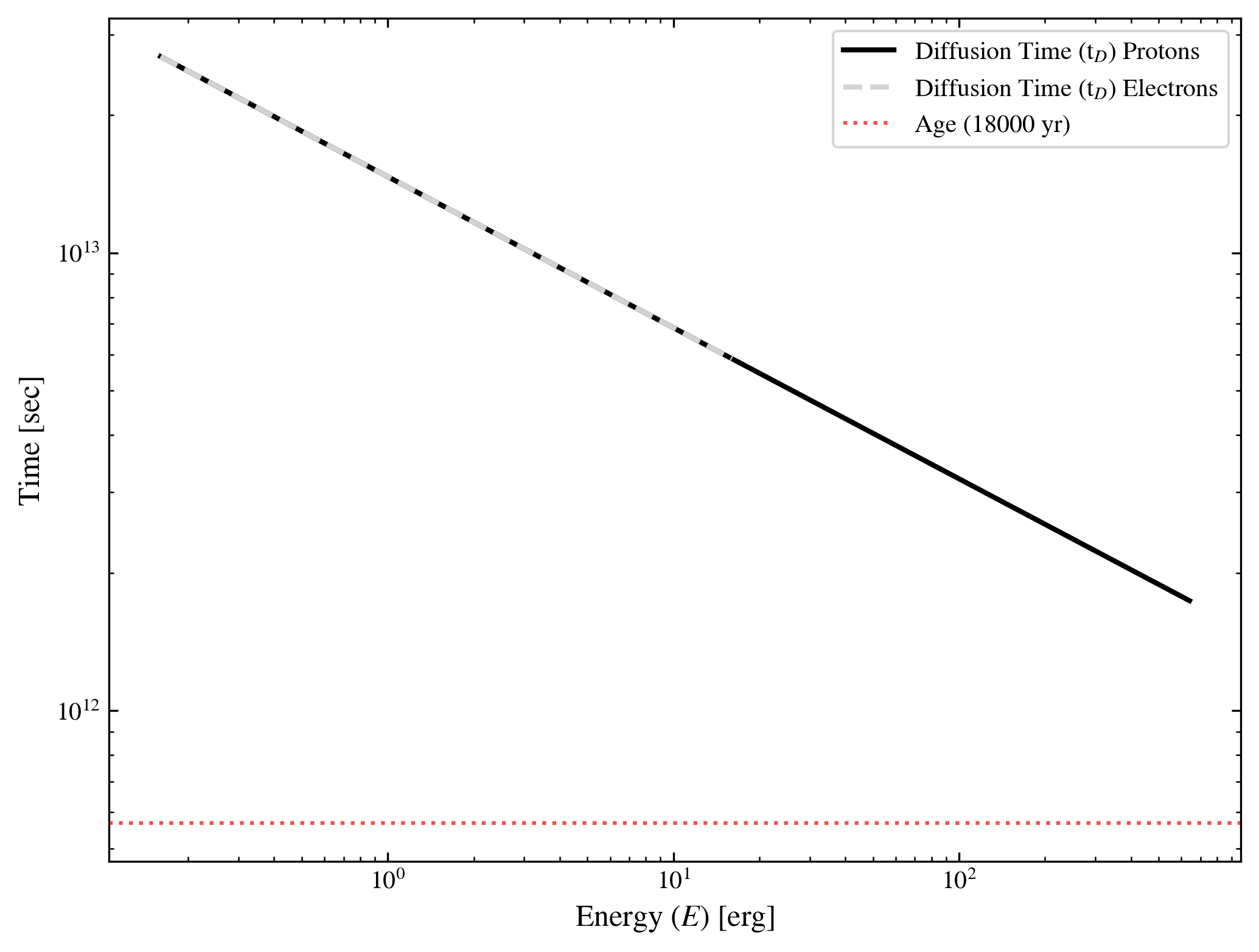}
    \label{fig:AgevsDiffTime}
\end{figure}

\section{MCMC consistency check of the main parameters used}
\label{App:B}

\begin{figure}[h!]
    \centering
    \includegraphics[width=0.5\linewidth]{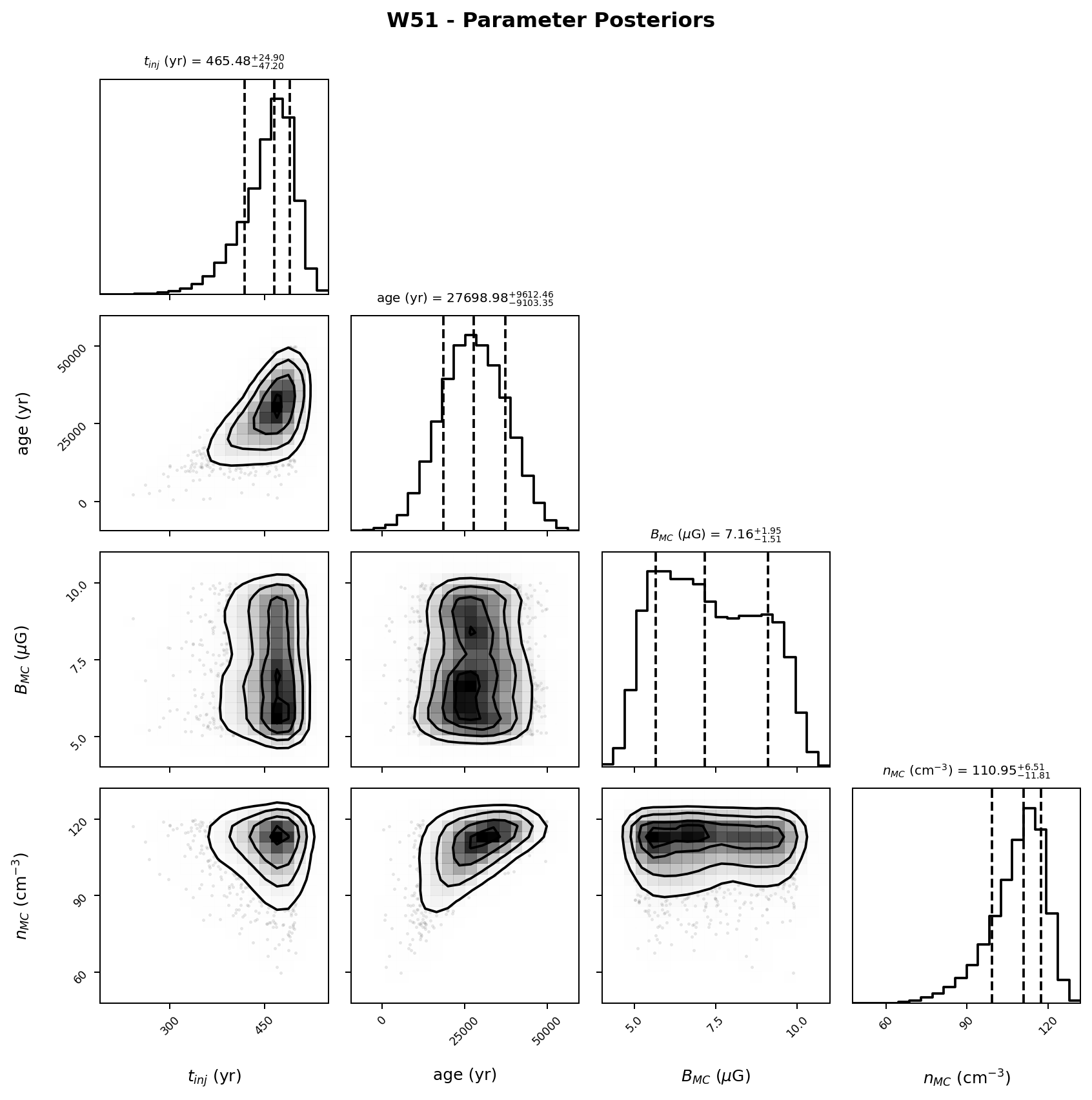}
    \caption{Parameter Posteriors  with a fixed injection spectrum into the MCs. The ranges well accommodate the values we obtain from our physically motivated fit. }
    \label{fig:cornerplot_mcmc}
\end{figure}

We performed a Bayesian Markov Chain Monte Carlo (MCMC) analysis pairing \textsc{GAMERA} with the emcee Python package \citep{Foreman_2013emcee}. The likelihood of the model was evaluated by calculating the $\chi^2$ statistic against the $\gamma$-ray spectral energy distribution measured by LHAASO. The important thing to note is that in our model setup, the spectral shapes and absolute luminosities of the injected particles were fixed as given above in Table. ~\ref{tab:Inj_Particle}. The total injected energy budget for each species therefore scales linearly with the free injection duration, $t_{inj}$. The particles were injected up to a maximum energy of $400,\mathrm{TeV}$ and subsequently underwent time-dependent cooling, diffusion, and multi-wavelength radiation. We allowed four key parameters to vary: the injection duration ($t_{\mathrm{inj}}$), the particle evolution time ($age$), the magnetic field in the MC ($B_{\mathrm{MC}}$), and the ambient density of the cloud ($n_{\mathrm{MC}}$). Uniform priors were adopted within physically motivated ranges: $age < 5 \times 10^{4}$ yr, $B_{\mathrm{MC}} < 10\,\mu\mathrm{G}$, $n_{\mathrm{MC}} < 120\,\mathrm{cm}^{-3}$, and $t_{\mathrm{inj}} < 500$ yr, considering particle confinement within the MC and short-period injection.

Since the full particle evolution calculation is computationally expensive, we used an emulator to speed up the MCMC sampling. We first generated a grid of model spectra covering the full parameter space, and then used interpolation to estimate the model output during the MCMC run. The analysis was performed with 32 walkers over 2000 steps, allowing an efficient exploration of the parameter space. The resulting parameter distributions are shown in Fig.~\ref{fig:cornerplot_mcmc}.

Owing to the computationally intensive nature of the underlying model and the coarse resolution of the emulator grid, the resulting chains are not intended to provide fully converged posterior distributions, but rather to verify that the parameter region identified through our physically motivated modeling is supported by the data. We find that the best-fit region broadly agrees with the parameters obtained from our model that gave the fit Fig. \ref{fig:Illum_SED}.

\end{appendix}
\FloatBarrier 
\clearpage

\end{document}